\documentstyle[prl,aps]{revtex}
\input BoxedEPS.tex
\SetepsfEPSFSpecial
\HideDisplacementBoxes

\begin{document}

\twocolumn[\hsize\textwidth\columnwidth\hsize
          \csname @twocolumnfalse\endcsname
\title{The magnetic resonance in high-temperature
superconductors:  Evidence for an extended s-wave
pairing symmetry}
\author{Guo-meng Zhao$^{*}$}
\address{ Department of
Physics and Astronomy, California State University at Los Angeles, Los
Angeles, CA 90032, USA}

\maketitle
\widetext
\begin{abstract}

We have identified several important features in the neutron 
scattering data of 
cuprates, which are difficult to be explained in terms of d-wave and 
isotropic s-wave 
order parameters.  Alternatively, we show that the neutron data are 
in 
quantitative agreement with an order parameter that has an extended 
s-wave ($A_{1g}$) symmetry and opposite sign in 
the bonding and antibonding electron bands formed within the 
Cu$_{2}$O$_{4}$ bilayers.  The extended s-wave has eight line nodes 
and change signs when a node is crossed.  This $A_{1g}$ pairing 
symmetry may be compatible with a charge fluctuation mediated pairing 
mechanism.

\end{abstract}
\narrowtext
\vspace{0.3cm}

]
The microscopic pairing mechanism responsible for high-temperature
superconductivity in copper-based perovskite oxides is still a 
subject of  intense debate despite tremendous experimental and 
theoretical efforts for over 15 years.  The debate has centered 
around 
the role of antiferromagnetic spin fluctuations in high-temperature 
superconductivity and the symmetry of superconducting condensate 
(order parameter).  Extensive inelastic neutron scattering 
experiments 
have accumulated a great deal of important data that should be 
sufficient to address these central issues.  Of particular interest 
is 
the magnetic resonance peak that has been observed in double-layer 
cuprate superconductors such as YBa$_{2}$Cu$_{3}$O$_{y}$ 
(YBCO) \cite{Bourges95,Fong,Bourges96,Bourges98,Bourges99,Dai} and 
Bi$_{2}$Sr$_{2}$CaCu$_{2}$O$_{8+y}$ (BSCCO) \cite{FongNature,He}, and 
in a 
single-layer compound Tl$_{2}$Ba$_{2}$CuO$_{6+y}$ (Tl-2201) 
\cite{He2002}.  A 
number of theoretical models \cite{Demler,Abanov,Morr,Brinckmann} 
have 
been proposed to explain the magnetic resonance peak in terms of  
d-wave 
magnetic pairing mechanisms.  These 
theories can qualitatively explain some features of neutron data but 
are 
particularly difficult to account for an important feature: The 
magnetic 
resonance in optimally and overdoped double-layer cuprates is much 
more 
pronounced in the odd channel than in the even channel 
\cite{Fong,Bourges96,Bourges98}.  In order to overcome this 
difficulty, Mazin and co-worker \cite{Mazin} proposed an order 
parameter that has isotropic s-wave symmetry and opposite sign in 
the 
bonding and antibonding electron bands formed within the 
Cu$_{2}$O$_{4}$ bilayers \cite{Mazin}.  However, this model predicts 
that the resonance energy is larger than twice the magnitude of the 
superconducting gap along the Cu-O bonding direction, in disagreement 
with experiment.  Further, the nodeless s-wave gap symmetry is 
inconsistent with the measurements of the penetration depth, thermal 
conductivity and specific heat, which consistently suggest the 
existence of line nodes in the gap function of hole-doped cuprates 
\cite{Hardy,Chiao}.

Here we identify several important features in the neutron scattering 
data of 
cuprates. These features are inconsistent with  d-wave order 
parameter (OP).  Alternatively, we show that the 
neutron data are in 
quantitative agreement with an order parameter that has an extended 
s-wave ($A_{1g}$) symmetry \cite{Zhao,Brandow} and opposite sign in 
the 
\begin{figure}[htb]	
    \ForceWidth{7cm}	
\centerline{\BoxedEPSF{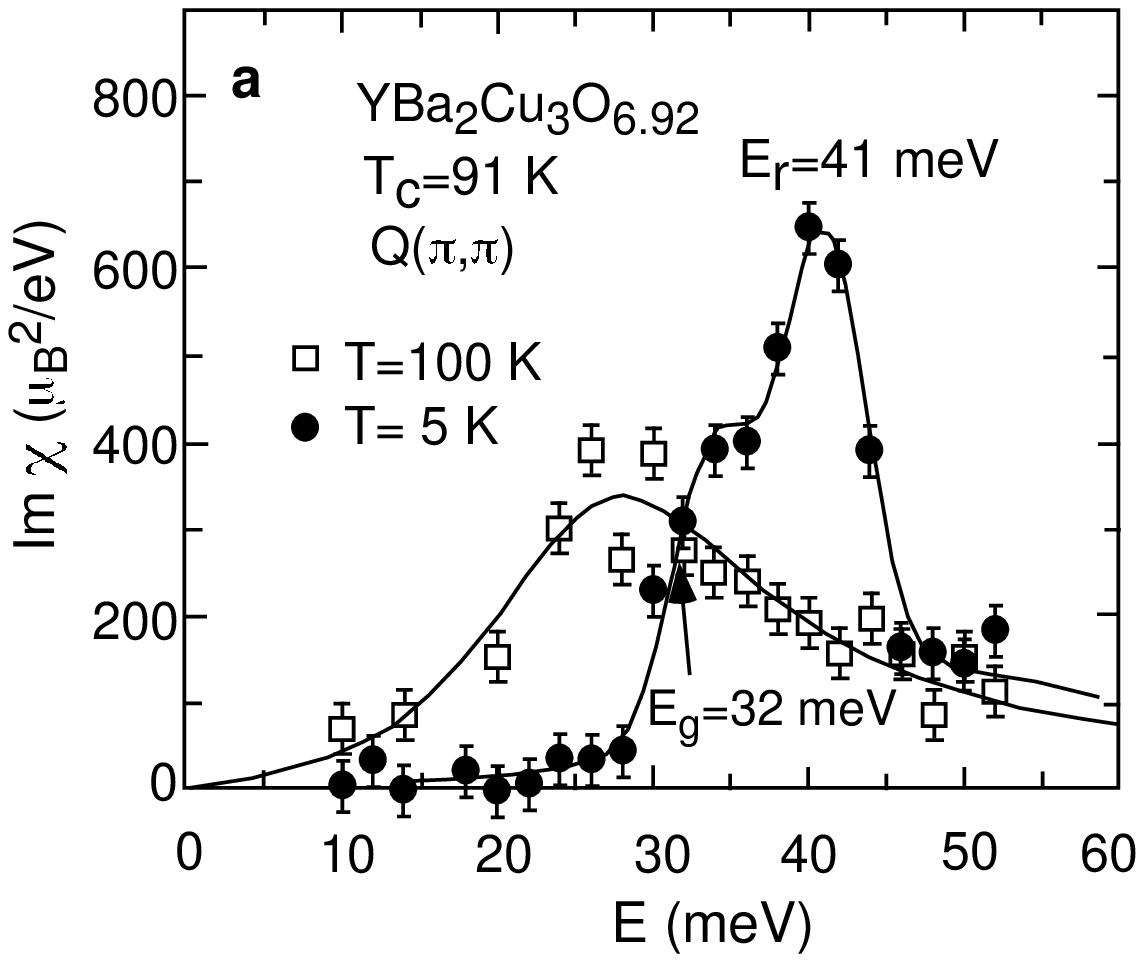}}	
\vspace{0.2cm}	
\ForceWidth{7cm}	
\centerline{\BoxedEPSF{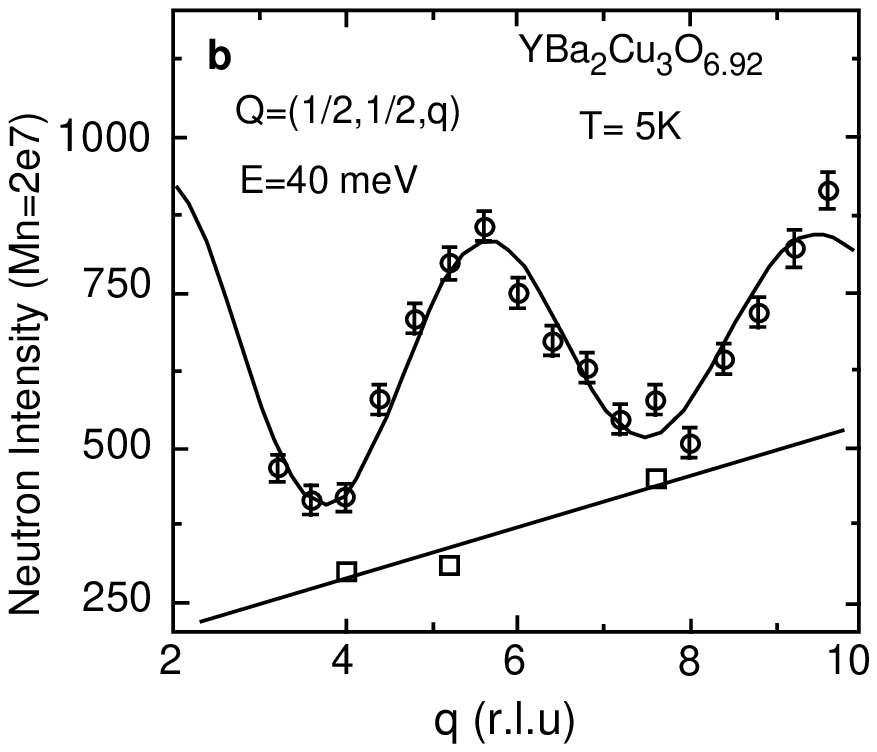}}	
	\vspace{0.3cm}
\caption[~]{a) The imaginary part of dynamic spin susceptibility as a 
function of excitation energy for YBa$_{2}$Cu$_{3}$O$_{6.92}$ in the 
normal and superconducting states.  The figure was reproduced from 
Ref.~\cite{Bourges99}.  b) $q_{l}$-scan at $E $= 40 meV in 
YBa$_{2}$Cu$_{3}$O$_{6.92}$ displaying a modulation typical of odd 
excitation.  The figure was reproduced from Ref.~\cite{Bourges98}.  
The background (lower line and open squares) was obtained from q-scans 
across the magnetic line.  The upper solid curve is a fit to 
$a+bF^{2}(\vec{Q})\sin^{2}(\pi zq_{l})$, where $F(\vec{Q})$ is the Cu 
magnetic form factor.  The modulation is not complete as even 
excitations are sizable at $q_{l}$= 3.5, 7 but with a magnitude 5 
times smaller \cite{Bourges98}.  }
\protect\label{Fig1}
\end{figure}
\noindent
bonding and antibonding electron bands formed within the 
Cu$_{2}$O$_{4}$ bilayers \cite{Mazin}.

 The magnetic resonance peak observed in optimally doped and 
overdoped double-layer compounds YBa$_{2}$Cu$_{3}$O$_{y}$  
\cite{Bourges95,Fong,Bourges96,Bourges98,Bourges99} and 
Bi$_{2}$Sr$_{2}$CaCu$_{2}$O$_{8+y}$  \cite{FongNature,He} is a 
sharp collective mode that occurs at an energy of 38-43 meV and at 
the 
two-dimensional wavevector $\vec{Q}_{AF}$ = ($\pi/a$, $\pi/a$), where 
$a$ is the nearest-neighbor Cu-Cu distance.  Fig.~1a shows the 
imaginary part of the odd channel spin susceptibility as a function 
of 
excitation energy for a slightly underdoped double-layer compound 
YBa$_{2}$Cu$_{3}$O$_{6.92}$.  The figure is reproduced from 
Ref.\cite{Bourges99}.  There are several striking features in the 
data: (a) The resonance peak at $E_{r}$= 41 meV appears below $T_{c}$ 
and this resonance peak intensity in the superconducting state 
$I_{odd}^{S}$($E_{r}$) is larger than the normal-state intensity 
$I_{odd}^{N}$($E_{r}$) by a factor of about 3.6, i.e., 
$I_{odd}^{S}(E_{r})$/$I_{odd}^{N}(E_{r})$ = 3.6; (b) A spin gap 
feature is seen below $T_{c}$ and there is a small shoulder that 
occurs at an energy slightly above the spin gap energy $E_{g}$ =32 
meV.  Fig.~1b shows a $q_{l}$-scan at $E $= 40 meV in 
YBa$_{2}$Cu$_{3}$O$_{6.92}$ displaying a modulation typical of odd 
excitation.  The figure is reproduced from Ref.\cite{Bourges98}.  
From 
Fig.~1b, we find feature (c): The odd-channel magnetic resonance 
intensity within the Cu$_{2}$O$_{4}$ bilayers is larger than the 
even-channel one by a factor of about 5 (Ref.\cite{Bourges98}), i.e., 
$I^{S}_{odd}(E_{r})/I^{S}_{even}(E_{r})$ = 5.  From the neutron 
studies on different double-layer compounds with different doping 
levels, one identifies feature (d): The resonance energy $E_{r}$ does 
not increase with doping in the overdoped range but is proportional 
to 
$T_{c}$ as $E_{r}/k_{B}T_{c}$ $\simeq$ 5.2 in both underdoped and overdoped 
ranges \cite{He}.

On the other hand, the neutron data for a single-layer Tl-2201 are 
different from those for double-layer compounds.  For the 
single-layer 
Tl-2201, we show in Fig.~2a the difference spectrum of the neutron 
intensities at $T$ = 27 K ($<$$T_{c}$) and $T$ = 99 K ($>$$T_{c}$), 
at 
a wavevector of $\vec{Q}$ = (0.5, 0.5, 12.25).  The figure is 
reproduced from Ref.~\cite{He2002}.  Although a sharp peak feature is 
clearly seen at $E_{r}$ = 46 meV, it is remarkable that the peak 
intensity in the superconducting state is close to  the magnetic 
neutron intensity in the normal state, that is,  
$I^{S}(E_{r})$/$I^{N}(E_{r})$$\simeq$ 1.  This is in sharp contrast 
with the above result for the 
double-layer compound YBa$_{2}$Cu$_{3}$O$_{6.92}$ where we found 
$I_{odd}^{S}(E_{r})$/$I_{odd}^{N}(E_{r})$ = 3.6.  Thus we identify 
feature (e) as $I^{S}(E_{r})$/$I^{N}(E_{r})$ $\simeq$ 1 for 
the single-layer compound.

We would like to mention that feature (e) identified for the 
single-layer Tl-2201 
is valid only if the nonmagnetic background has a negligible 
temperature dependence below 100 K.  It is known that nuclear 
contributions predominantly constitute the main part of the 
neutron signal \cite{Bourges96}.  Any nuclear contributions 
(nonmagnetic backgrounds) are only 
expected to obey the following standard temperature dependence, 
$1/[1-\exp(-E/k_{B}T)]$ (see detailed discussions in 
Ref.~\cite{Bourges96}).  
In the energy range of interest ($E$ $>$ 40 meV), this temperature 
dependence factor is close to unity and nearly independent of 
temperature for $T < $ 150 K.  Indeed, the normal-state neutron 
intensities of both 
YBCO \cite{Bourges96} and BSCCO \cite{FongNature,He} show a 
negligible 
temperature dependence for $T_{c}$ $<$ $T$ $<$ 150 K.  For ~~$E$ = 
\begin{figure}[htb]	
    \ForceWidth{7cm}	
\centerline{\BoxedEPSF{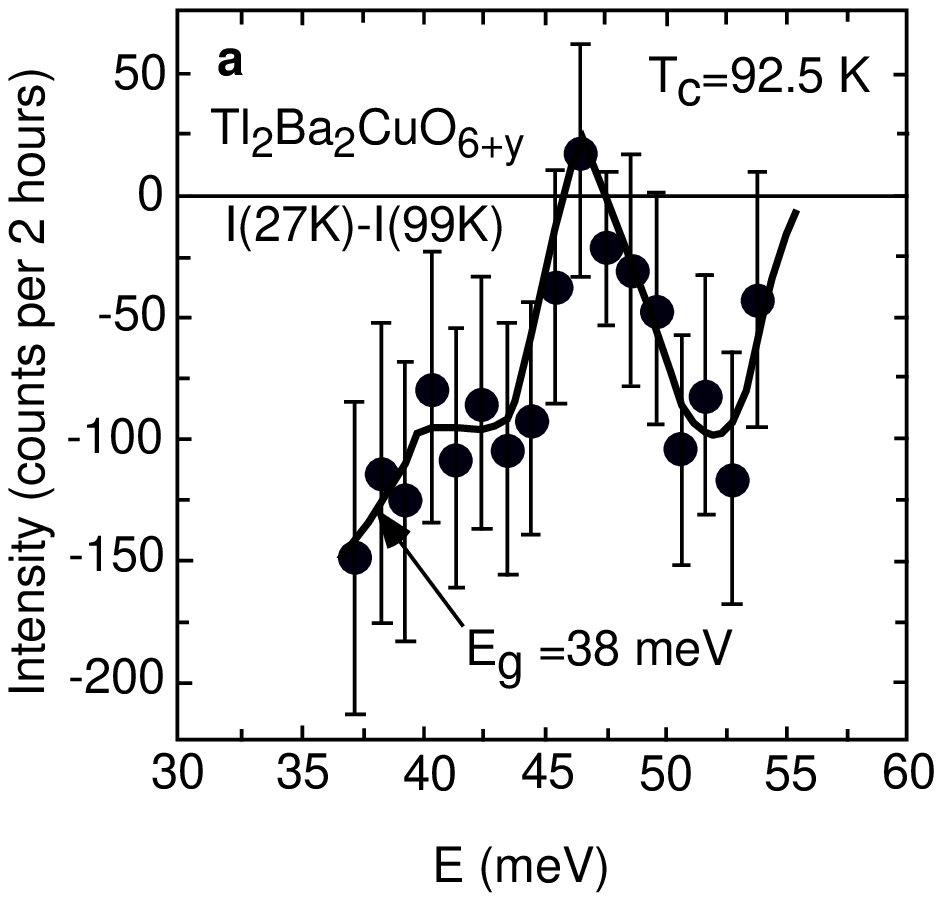}}
\ForceWidth{7cm}
\vspace{0.3cm}	
\centerline{\BoxedEPSF{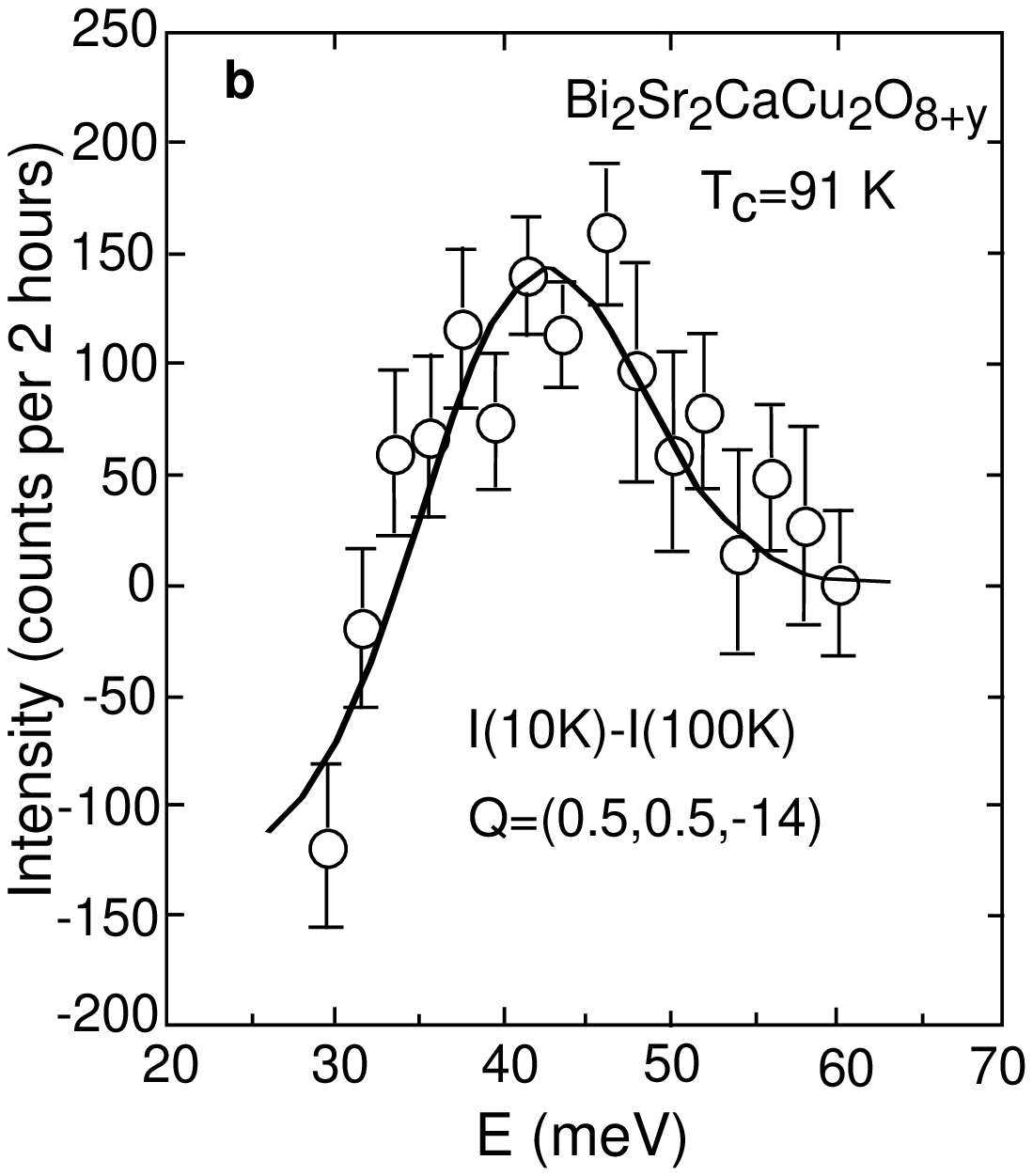}}	
	\vspace{0.3cm}
\caption[~]{a) The difference spectrum of the neutron intensities 
of single-layer Tl-2201 crystals (with a total volume of 0.11 
cm$^{3}$) at 
$T$ =27 K ($<$$T_{c}$) and $T$ = 99 K ($>$$T_{c}$), and at wavevector 
$\vec{Q}$ = (0.5, 0.5, 12.25).  The figure is reproduced from 
Ref.~\cite{He2002}.  The solid line is guide to the eye. The 
difference spectrum tends to zero at about 56 meV (about 10 meV 
higher than the resonance energy). This  suggests that the 
nonmagnetic background at 56 meV has negligible temperature 
dependence below 100 K.  b) The difference spectrum of the neutron 
intensities of optimally doped  BSCCO crystals (with a total volume 
of 0.06 
cm$^{3}$) at $T$ =10 K ($<$$T_{c}$) and $T$ = 100 K ($>$$T_{c}$), and 
at wavevector $\vec{Q}$ = (0.5, 0.5, -14).  The figure is reproduced 
from Ref.~\cite{FongNature}. The 
difference spectrum goes to zero at 60 meV (about 16 meV 
higher than the resonance energy). This  suggests that the 
nonmagnetic background at 60 meV has negligible temperature 
dependence below 100 K.    }
\protect\label{Fig2}
\end{figure}
\noindent
$E_{r}$ = 46 meV in Tl-2201, the factor $1/[1-\exp(-E/k_{B}T)]$ 
decreases by 0.4$\%$ when 
the temperature is lowered  from 99 K to 27 K.  If we take the 
nonmagnetic 
background of 2500 counts per two hours \cite{He2002}, the 
nonmagnetic 
background decreases by 10 counts per two hours when the temperature 
goes from 99 K to 27 K, and by 60 counts per two hours when the 
temperature goes from 150 K to 99 K.  The variation of the 
nonmagnetic 
background at $E$ = 46 meV below 100 K is negligibly small compared 
with the magnetic resonance intensity ($\sim$150 counts per two 
hours).  Therefore, feature (e) identified for the single-layer 
Tl-2201 is well justified.

In order to further justify feature (e), we compare the ratios of the 
magnetic to the nonmagnetic background signals in those neutron 
experiments on different compounds.  From the neutron data, we find 
that the 
signal-to-background ratio is about 6$\%$ for Tl-2201
(Ref.~\cite{He2002}), about 12$\%$ for an
optimally doped
BSCCO \cite{FongNature}, about 6$\%$ for an overdoped BSCCO 
\cite{He}, and about 100$\%$ for an overdoped YBCO \cite{Bourges96}. 
The
signal-background-ratio in the overdoped BSCCO is similar to  that 
for 
Tl-2201.  But the overdoped BSCCO shows a difference spectrum similar 
to that for the overdoped YBCO where the signal-background-ratio is 
larger than that for the overdoped BSCCO by a factor of 20.  This 
suggests that, if feature (e) identified for Tl-2201 were an artifact 
caused by a small signal-background-ratio, one would not have 
observed 
the intrinsic difference spectra for the overdoped BSCCO.  The fact that the 
difference spectra for the overdoped BSCCO are similar to the one for YBCO suggests 
that one indeed finds the intrinsic magnetic difference spectra for 
the overdoped BSCCO even though the signal-background-ratio is only 6$\%$.  There 
are no
reasons to
believe 
that only the difference spectrum for Tl-2201 is an artifact.  
Therefore, the 
pronounced difference between the difference spectrum of the 
single-layer Tl-2201 and that of the double-layer BSCCO (compare Fig.~2a 
and Fig.~2b) is due to the fact that 
$I_{odd}^{S}(E_{r})$/$I_{odd}^{N}(E_{r})$ $>$$>$ 1 for the 
double-layer BSCCO while $I^{S}(E_{r})$/$I^{N}(E_{r})$ $\sim$ 1 for 
the 
single-layer Tl-2201.  Moreover, we will show below that, for the 
intraband scattering (even channel) in YBa$_{2}$Cu$_{3}$O$_{6.92}$, 
$I_{even}^{S}(E_{r})$/$I_{even}^{N}(E_{r})$ = 0.72, in good agreement 
with feature (e) for the single-layer Tl-2201.  This consistency 
gives 
further support to the thesis that feature (e) is intrinsic.

One may also argue that the glue that glues about 300 small 
crystals of Tl-2201 on Al-plates would cause a  substantial  decrease 
of  the nonmagnetic 
background  
below 100 K. If this argument were relevant,  one would have observed 
a similar effect in overdoped  
Y$_{0.9}$Ca$_{0.1}$Ba$_{2}$Cu$_{3}$O$_{7-y}$ (YBCO-Ca) because 
60 larger crystals of YBCO-Ca are also glued on 
Al-plates \cite{Pai}.  Because the total volume of YBCO-Ca crystals 
is 
larger than that of Tl-2201 crystals by a factor of 3.2, one can 
readily show 
that the amount of glue for YBCO-Ca crystals is comparable with or 
even larger than that for Tl-2201.  This suggests that the effect of 
glue on the nonmagnetic background in Tl-2201 is similar to that in 
YBCO-Ca.  From the data for YBCO-Ca (Fig.~2a and Fig.~3a of 
Ref.~\cite{Pai}), one can clearly see that the difference spectrum at 
50 meV is very close to zero.  Because the magnetic signal at an 
energy that is about 10 meV higher than the resonance energy is 
independent of temperature below 100 K (see Fig.~1a), the nearly zero 
value of the difference spectrum at 50 meV implies that the nonmagnetic 
background at 50 meV has negligible temperature dependence below 100 K 
in the case of YBCO-Ca.  Even in the case of Tl-2201 (see Fig.~2a ), 
the difference spectrum tends to zero at about 56 meV (about 10 meV 
higher than the resonance energy) although one needs more data points 
in the vicinity of 56 meV to definitively address this issue.  This 
indicates that the nonmagnetic background at 56 meV has negligible 
temperature dependence below 100 K in the case of Tl-2201.

We would like to mention that  the normal-state magnetic 
intensities in Fig.~4 of Ref.~\cite{He2002} are significantly 
underestimated.  
This is because the authors assume that the $q$-width of the 
magnetic peak 
in the normal state is the same as that in the superconducting state 
\cite{He2002}.  This assumption is unphysical.  From Fig.~19d and 
Fig.~19h 
of Ref.~\cite{Dai}, one can clearly see that the $q$-width of the 
magnetic peak in the normal state is a factor of 1.6  larger than 
that in the superconducting state in the case of underdoped 
YBa$_{2}$Cu$_{3}$O$_{6.8}$.  Further, the $q$-width increases with 
the increase of doping.  With a much broader magnetic peak in the normal 
state, the $q$ range (0.35 rlu - 0.65 rlu) for the normal-state 
$q$-scan spectrum (see Fig.~2B of Ref.\cite{He2002}) is too narrow to get meaningful estimates 
of the nonmagnetic background and the normal-state magnetic intensity.
Moreover, the normal-state $q$-scan spectrum of 
YBa$_{2}$Cu$_{3}$O$_{6.8}$ is 
nearly featureless for the same narrow $q$-range (0.35 rlu - 0.65 rlu), 
similar to the data of Fig.~2B of Ref.~\cite{He2002}.  If one would fit the 
normal-state $q$-scan data only in this narrow $q$-range (0.35 rlu - 0.65 
rlu) for YBa$_{2}$Cu$_{3}$O$_{6.8}$ and assume the same $q$-width as 
that in the superconducting state, one would find that 
the normal-state magnetic intensity is underestimated by a factor of 
4.  Such a significant underestimate may be also true for the 
normal-state magnetic intensities of Tl-2201.  The authors of 
Ref.~\cite{He2002} should have extended their measurements to a wider 
$q$ range to get reliable estimates of the nonmagnetic background and 
the normal-state magnetic intensities.

Since magnetic signals in both normal and 
superconducting states strongly depend on doping, as clearly seen in 
YBCO \cite{Bourges98}, the comparison of the resonance peak 
intensities and spectral weights among different compounds should be 
carefully made.  For this Tl-2201, there is some indication of a broad 
peak centered around $\vec{Q}_{AF}$ even above T$_{c}$ (Ref.  
\cite{He2002}), as observed in underdoped YBCO \cite{Bourges98,Dai}.  
This suggests that the Tl-2201 is slightly underdoped so that the 
magnetic intensity in the normal state should be close to that for 
slightly underdoped YBa$_{2}$Cu$_{3}$O$_{6.92}$.  This is because both 
compounds have a similar ratio T$_{c}$/T$_{cm}$ and thus a similar 
doping level (where T$_{cm}$ is the superconducting transition at 
optimal doping).  As seen from Fig.~1a, the normal-state magnetic 
intensity at 40 meV is about 100 $\mu_{B}^{2}$/eV per Cu in slightly 
underdoped YBa$_{2}$Cu$_{3}$O$_{6.92}$.  The resonance peak intensity 
for the Tl-2201 can be estimated to be about 120 $\mu_{B}^{2}$/eV per 
Cu from the resonance spectral weight of 0.7$\mu_{B}^{2}$ per Cu 
(Ref.~\cite{He2002}).  Therefore, the resonance peak intensity for the 
Tl-2201 is close to the normal-state magnetic intensity for the 
slightly underdoped YBa$_{2}$Cu$_{3}$O$_{6.92}$. Since the normal-state magnetic intensity 
for Tl-2201 should be similar to 
that for  the 
slightly underdoped YBa$_{2}$Cu$_{3}$O$_{6.92}$, then $I^{S}(E_{r})$/$I^{N}(E_{r})$ $\simeq$ 1.2 for this 
single-layer Tl-2201.  This is in good agreement with feature (e) 
deduced independently from the difference spectrum (Fig.~2a).  
Moreover, we find that the resonance peak intensity for the Tl-2201 
($\sim$120 $\mu_{B}^{2}$/eV per Cu) is a factor of about three smaller 
than the resonance peak intensity for the slightly underdoped 
YBa$_{2}$Cu$_{3}$O$_{6.92}$ ($\sim$330 $\mu_{B}^{2}$/eV per Cu).  
Similarly, the resonance spectral weight for the Tl-2201 is also a 
factor of three smaller than that for the slightly underdoped 
YBa$_{2}$Cu$_{3}$O$_{6.92}$.

These important features we have identified above should place strong 
constraints on theories for high-temperature 
superconductivity in cuprates.  Any correct theories should be able 
to  
explain all the magnetic resonance features in a consistent and 
quantitative way.  
In a more exotic approach \cite{Demler}, the 
neutron data are interpreted in terms of a collective mode in the 
spin-triplet particle-particle channel, which couples to the 
particle-hole channel in the superconducting state with d-wave OP.  
This model predicts that the 
resonance peak energy $E_{r}$ is proportional to the doping level 
$p$, in disagreement with feature (d):  $E_{r}$ does not increase 
with 
increasing $p$ in the overdoped range but is proportional to $T_{c}$ 
as $E_{r}/k_{B}T_{c}$ $\simeq$ 5.2 (Ref.~\cite{He}).  Moreover, this model  predicts \cite{Norman}  
that $E_{r}  > 2\Delta_{M}$ (where $\Delta_{M}$ is the maximum d-wave 
gap), 
which contradicts experiment. 
Other theories based on spin-fermion interactions also show 
that $E_{r}$ increases monotonically with increasing $p$ 
\cite{Abanov,Morr},  in disagreement with  feature (d).

Alternatively, feature (d)  is consistent with a simple particle-hole 
excitation across the  
superconducting gap within an itinerant magnetism model. This is 
because the particle-hole excitation 
energy increases with the superconducting gap which in turn should be 
proportional to  $T_{c}$ 
at least in the overdoped range.  This itinerant magnetism model is 
also 
supported by  very 
recent Fourier transform scanning
tunnelling spectroscopic (FT-STS) studies on a nearly optimally doped 
BSCCO
\cite{Davis}, which show that the quasiparticles in the 
superconducting state exhibit particle-hole mixing similar to that of 
conventional Fermi-liquid superconductors.  These FT-STS results thus 
provide evidence for a Fermi-liquid behavior  in the superconducting 
state of optimally doped and overdoped cuprates.  Here we 
quantitatively 
explain all these neutron data 
\cite{Bourges95,Fong,Bourges96,Bourges98,Bourges99,Dai,FongNature,He,He2002} 
in terms of an order parameter (OP) that has an extended s-wave 
symmetry \cite{Zhao,Brandow} and opposite sign in the bonding and 
antibonding electron bands formed within the Cu$_{2}$O$_{4}$ bilayers 
\cite{Mazin}.  In our model, the neutron resonance peak is due to 
excitations of electrons from the extended saddle points below the 
Fermi level to the superconducting gap edge above the Fermi level.

Within the itinerant magnetism model, neutron scattering intensity at 
a wavevector
$\vec{q}$ and an energy $E$ is proportional to the imaginary part of 
the dynamic
electron spin susceptibility, $\chi^{\prime\prime}(\vec{q}, E)$. 
Qualitatively,
with the neglect of the Bardeen-Cooper-Schrieffer (BCS) coherence 
factor, 
the bare imaginary part of spin susceptibility, 
$\chi^{\prime\prime}_{\circ}(\vec{q}, E)$, is proportional to the 
joint density of states $A(\vec{q}, E)= \sum_{\vec{k}}{\delta (E - 
E_{\vec{k}+\vec{q}}-E_{\vec{k}})}$, where $E_{\vec{k}} 
=\sqrt{\epsilon_{\vec{k}}^{2}+\Delta^{2}_{\vec{k}}}$ is the 
quasiparticle dispersion law below $T_{c}$, $\epsilon_{\vec{k}}$ is 
the electronic band dispersion, and $\Delta_{\vec{k}}$ is the order 
parameter \cite{Mazin}.  The two particle energy 
$E_{2}(\vec{k},\vec{q})=E_{\vec{k}+\vec{q}}+E_{\vec{k}}$ has a 
minimum 
and several saddle points for a fixed $\vec{q}$.  The minimum defines 
the threshold energy, or spin gap energy $E_{g}$, which is achieved 
at 
vector $\vec{k}$ and $\vec{k}+\vec{q}$ such that both $\vec{k}$ and 
$\vec{k}+\vec{q}$ belong to the Fermi surface and thus $E_{g} = 
2\Delta_{\vec{k}}$ (Ref.~\cite{Mazin}).

The joint density of
states has  divergences at the extended saddle
points \cite{Mazin}. The divergent peak in 
$\chi^{\prime\prime}_{\circ}(\vec{q},
\omega)$ occurs because of transitions between the occupied states
located in the extended saddle points below $E_{F}$ and empty
quasiparticle states at the superconducting gap edge above $E_{F}$. 
The saddle points
produce Van Hove singularities  in
the quasiparticle density of states in the superconducting state at
an energy of 
$\sqrt{(\epsilon^{VH}_{\vec{k}})^{2}+\Delta_{\vec{k}}^{2}}$
and the superconducting condensate creates a sharp coherence peak in
the quasiparticle density of states at
the gap edge.  Thus, the
divergence in the joint density of states in the superconducting
state is then located at the energy
$E^{*} 
=\Delta_{\vec{k}+\vec{q}}+\sqrt{(\epsilon^{VH}_{\vec{k}})^{2}+\Delta_{\vec{k}}^{2}}$.  This simple expression for $E^{*}$ has been verified by numerical 
calculations in the case of an isotropic s-wave order parameter  
\cite{Mazin}.

In Fig.~3,  we plot the Fermi surface for a  slightly underdoped 
BSCCO,	
which is inferred by extending a portion of the Fermi surface 
determined by	
angle-resolved photoemission spectroscopy (ARPES) \cite{White}.  There
are extended saddle points that are located near ($\pm$$\pi$, 0) and 
(0, $\pm$$\pi$), as shown by ARPES \cite{Dessau}.  A solid line in 
Fig.~3
represents a segment of the extended saddle points very close to the 
Fermi surface. 
Other portions of  the saddle points [e.g., along the line from (0, 
0) 
to ($\pi$, 0) ] are significantly away from the 
Fermi surface \cite{Dessau} and not shown in the figure.  If we only 
consider the
magnetic excitations at a fixed wavevector that corresponds to the
antiferromagnetic wavevector $\vec{Q}_{AF}$, only four electron
wavevectors at the Fermi surface are connected by $\vec{Q}_{AF}$ as
indicated by arrow 1 in Fig.~3.  Each of these vectors forms an angle
of $\theta_{r}$ with respect to the Cu-O bonding direction.  The
electron transitions from the occupied states located at the extended
saddle points below $E_{F}$ to empty quasiparticle states at the
superconducting gap edge above $E_{F}$ are indicated by arrow 2.
Because the quasiparticle densities of states at the gap edge and the
saddle points are divergent at zero temperature, such transitions will
produce a sharp resonance peak at $E_{r}
=\Delta_{\vec{k}+\vec{Q}_{AF}}+\sqrt{(\epsilon^{VH}_{\vec{k}})^{2}+\Delta_{\vec{k}}^{2}}$  (Ref.~\cite{Mazin}). 

\begin{figure}[htb]
    \ForceWidth{7cm}
\centerline{\BoxedEPSF{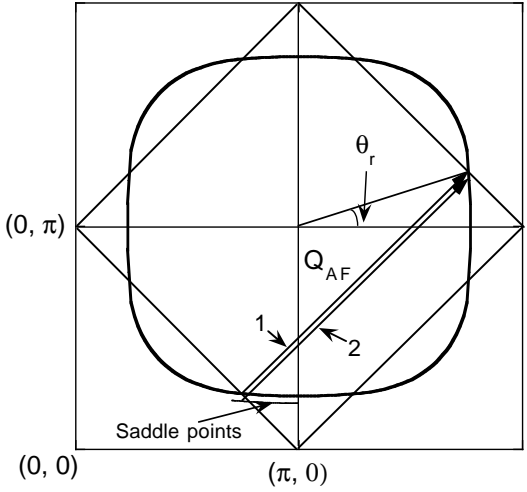}}	
	\vspace{0.3cm}
	\caption[~]{The Fermi surface for a slightly underdoped BSCCO
with $T_{c}$ = 88 K.  This Fermi-surface is extrapolated from a part of the Fermi	
surface determined by ARPES \cite{White} using symmetry arguments.  Arrow 1	
indicates electron transitions from the occupied states in the	
superconducting gap edge below $E_{F}$ to empty quasiparticle states	
at the gap edge above $E_{F}$.  Arrow 2 marks electron transitions	
from the occupied states located in the extended saddle points below	
$E_{F}$ to empty quasiparticle states in the superconducting gap edge	
above $E_{F}$.  }	
\protect\label{Fig3}	
\end{figure}
In terms of $\theta_{r}$, both $E_{g}(\vec{Q}_{AF})$ and
$E_{r}(\vec{Q}_{AF})$ can be rewritten as
\begin{equation}\label{Ne1}
E_{g} (\vec{Q}_{AF})= 2\Delta (\theta_{r})
\end{equation}
and
\begin{equation}\label{Ne2}
E_{r} (\vec{Q}_{AF})= \Delta (\theta_{r}) + \sqrt{[\Delta 
(\theta_{r})]^{2} +
(\epsilon^{VH}_{r})^{2}}.
\end{equation}
Here   $\sqrt{[\Delta (\theta_{r})]^{2} +(\epsilon^{VH}_{r})^{2}}$ is 
the energy of a saddle point below the 
Fermi
level along the $\theta_{r}$ direction.  One should note that
Eq.~\ref{Ne2} is valid only if the saddle points are very close to the
Fermi surface, as is the case.

We now consider the BCS coherence factor that has been ignored in the 
above discussions. The BCS coherence factor is given by \cite{Book}:
\begin{equation}\label{coh1}
\xi (\vec{q},\vec{k})=\frac{1}{2}(1-\frac{\epsilon_{\vec{k}}
\epsilon_{\vec{k}+\vec{q}}+\Delta_{\vec{k}+\vec{q}}\Delta_{\vec{k}}}
{E_{\vec{k}}E_{\vec{k}+\vec{q}}}).
\end{equation}
For $\epsilon_{\vec{k}}$$<$$<$ $\Delta_{\vec{k}}$
and $\epsilon_{\vec{k}+\vec{q}}$ $<$$<$ $\Delta_{\vec{k}+\vec{q}}$, 
the
BCS coherence factor is close to 1 when $\Delta_{\vec{k}+\vec{q}}$ and
$\Delta_{\vec{k}}$ have opposite sign, but is close to zero when
$\Delta_{\vec{k}+\vec{q}}$ and $\Delta_{\vec{k}}$ have the same sign.
The coherence factor is not negligible for
$\epsilon_{\vec{k}}$ $\simeq$ $\Delta_{\vec{k}}$ and
$\epsilon_{\vec{k}+\vec{q}}$ = 0 even if $\Delta_{\vec{k}+\vec{q}}$
and $\Delta_{\vec{k}}$ have the same sign.

For a single-layer compound
with d-wave order parameter symmetry,  $\Delta_{\vec{k}+\vec{Q}_{AF}}$
and $\Delta_{\vec{k}}$ have opposite sign so that
the BCS coherence factor $\simeq$ 1 and thus 
$I^{S}(E_{r})$/$I^{N}(E_{r})$$>$$>$ 1 (see Ref.~\cite{Brinckmann}).  
Therefore, the experimental observation of 
$I^{S}(E_{r})$/$I^{N}(E_{r})$ $\sim$ 1.0 in the single-layer Tl-2201 
\cite{He2002} rules out the d-wave order parameter.  Alternatively, 
for a single-layer compound with an s-wave symmetry, 
$\Delta_{\vec{k}+\vec{Q}_{AF}}$ and $\Delta_{\vec{k}}$ have the same 
sign, so the BCS coherence factor could be much less than 1.  This may 
lead to $I^{S}(E_{r})$/$I^{N}(E_{r})$$\sim$ 1, in agreement with feature (e).  Hence, only the intralayer (intraband) s-wave symmetry is 
compatible with feature (e): $I^{S}(E_{r})$/$I^{N}(E_{r})$ $\sim$ 
1.0.  On the other hand, if extended saddle points are significantly 
below the Fermi level, the BCS coherence factor will be substantial 
even for the intralayer (intraband) magnetic scattering in the case of 
s-wave gap symmetry.  Then $I^{S}(E_{r})$/$I^{N}(E_{r})$ could be much 
larger than 1 in this special case.  Therefore, the observation of 
$I^{S}(E_{r})$/$I^{N}(E_{r})$ $>>$ 1 for the intralayer (intraband) 
magnetic scattering is consistent with either s-wave or d-wave gap 
symmetry.

For a double-layer compound, interactions within Cu$_{2}$O$_{4}$ 
bilayers yield bonding
and antibonding bands. Transitions between electronic states
of the same type (bonding-to-bonding or antibonding-to-antibonding) 
and those of
opposite types are characterized by even or odd symmetry, 
respectively, under exchange
of two adjacent CuO$_{2}$ layers.  As a result, the magnetic
excitations between different bands correspond to odd channel 
excitations 
while the excitations within the same band to even channel 
excitations 
\cite{Bourges98}.  If the order parameters in the bonding and 
antibonding bands have the same sign, the magnetic resonance 
intensities in both channels should be similar for an intraband 
pairing symmetry of either s-wave or 
d-wave.  This is because the BCS 
coherence factors for both channels are the same in this case.  On 
the 
other hand, if the order parameter has an s-wave symmetry and 
opposite 
sign in the bonding and antibonding electron bands, the magnetic 
resonance intensity in the odd channel will be much larger than that 
in the even channel due to the large difference in their BCS 
coherence 
factors.  Therefore, only if the order parameter has an s-wave 
symmetry and opposite sign in the bonding and antibonding electron 
bands, can the observed feature (c): 
$I^{S}_{odd}(E_{r})/I^{S}_{even}(E_{r})$ = 5 for the slightly 
underdoped  YBCO and $I^{S}_{odd}(E_{r})/I^{S}_{even}(E_{r})$ $>$10 
for the overdoped YBCO \cite{Bourges96} be explained within the 
itinerant magnetism approach.

Based on the t-J model and d-wave pairing symmetry, Brinckmann and 
Lee \cite{Lee} have recently 
attempted to account for  feature (c) using a very unrealistic 
parameter:  $J_{\perp}$/$J_{\parallel}$ = 0.6 
(where $J_{\parallel}$ and $J_{\perp}$ are the effective 
intralayer and interlayer antiferromagnetic exchange energies, 
respectively). For undoped YBCO, neutron experiments \cite{Bourges98} 
show that 
$J_{\perp}$/$J_{\parallel}$ = 0.1, which is far less than 0.6 used 
in the calculation \cite{Lee}. Further,  the value of 
$J_{\perp}$/$J_{\parallel}$ will be negligible 
if $J_{\parallel}$ remains substantial and the optical magnon gap 
($\propto$ $\sqrt{J_{\parallel}J_{\perp}}$) goes to zero, which 
should be the case 
for optimally doped and overdoped YBCO.  Neutron 
data for underdoped YBCO \cite{Fong2000} show that the optical 
magnon gap is about 50 meV for YBa$_{2}$Cu$_{3}$O$_{6.5}$ and is 
reduced to about 25 meV for YBa$_{2}$Cu$_{3}$O$_{6.7}$.   If we 
linearly extrapolate the optical magnon gap 
with the oxygen content, the gap will tend to zero in 
YBa$_{2}$Cu$_{3}$O$_{y}$ 
for $y$ $>$ 0.9, implying that $J_{\perp}$/$J_{\parallel}$ $<<$ 0.1 
for 
optimally doped and overdoped cuprates. 

Using a more realistic parameter of $J_{\perp}$ and the d-wave 
pairing symmetry, Millis and 
Monien \cite{Millis} appear to be able to explain feature (c).  They 
showed that the resonance spectral weights in the odd 
and even channels are related to the difference between the resonance 
energy $E_{r}$ and the particle-hole spin excitation energy 2$\Delta 
(\theta_{r}$) at $\vec{Q}_{AF}$, i.e., $W^{odd}/W^{even}$ = $[2\Delta 
(\theta_{r})-E_{r}^{odd}]/[2\Delta (\theta_{r})-E_{r}^{even}]$.  Here 
2$\Delta (\theta_{r})$ must be larger than both $E_{r}^{even}$ and 
$E_{r}^{odd}$ (Ref.\cite{Millis}).  The INS data of 
Y$_{0.9}$Ca$_{0.1}$Ba$_{2}$Cu$_{3}$O$_{7-y}$ would be consistent with 
this 
theoretical model if one would choose an unrealistic parameter 
2$\Delta 
(\theta_{r})$ = 49 meV (Ref.~\cite{Pai}). As discussed below, 
$\theta_{r}$ 
is about 15$^{\circ}$ for slightly overdoped cuprates so that 
2$\Delta 
(\theta_{r})$ =1.73$\Delta_{M}$ (where $\Delta_{M}$ is the maximum 
d-wave gap).  The intrinsic tunneling spectra, which are  
unsusceptible
to surface deterioration, can provide the most reliable 
determination of the bulk superconducting gap \cite{Kras}. From the 
intrinsic tunneling spectra, one finds that 2$\Delta_{M}/k_{B}T_{c}$ 
= 
8.09 for the optimally doped BSCCO with T$_{c}$ = 94 K 
(Ref.~\cite{Kras}),  2$\Delta_{M}/k_{B}T_{c}$ = 6.73 for a slightly overdoped BSCCO 
with 
\widetext
\begin{table}[htb]
 \caption[~]{Comparison of experiments with extended s-wave, 
d-wave, and isotopic s-wave.  In both extended and isotropic s-wave 
models, the order parameters are assumed to have opposite signs in the 
bonding and antibonding electron bands formed within the 
Cu$_{2}$O$_{4}$ bilayers.  Here DA = definitive agreement, A = 
agreement, QA = qualitative agreement, PA = possible agreement, D = 
disagreement and DD = definitive disagreement.}
\begin{center}
	\begin{tabular}{llll}
INS data &extended s-wave&d-wave&isotropic s-wave\\
\hline
Feature (a): $I^{S}$($E_{r}$)/$I^{N}$($E_{r}$) = 3.6 for YBCO & DA& 
DA & DA\\
Feature (c):  $I^{S}_{odd}(E_{r})/I^{S}_{even}(E_{r})$ = 5 for YBCO & 
DA& DD & DA\\
Feature (e): $I^{S}$($E_{r}$)/$I^{N}$($E_{r}$) $<$1 for Tl-2201 & DA& DD & DA\\
The magnitudes of $E_{r}$ and $E_{g}$ & DA& DD & DD\\
Spin gap in La$_{2-x}$Sr$_{x}$CuO$_{4}$	& DA&		DD	&	DD\\	
\hline
Other data &extended s-wave&d-wave&isotropic s-wave\\
\hline
Penetration depth	&DA&QA&DD\\
Thermal conductivity	&	DA&		QA&DD\\
Specific heat			&DA&		QA&	DD\\
Nonlinear Meissner effect	&DA&	QA &		DD\\
ARPES (underdoped)&			A	&		A&	DD\\
ARPES (overdoped)		&	DA&	DD&	DD\\
Quasiparticle tunneling (underdoped)&	PA&		PA&	DD\\
Quasiparticle tunneling (overdoped) &	DA&		DD&	DD\\
Raman scattering (underdoped)&	PA	&		PA	&	DD\\
Raman scattering (overdoped)&DA&		DD&		PA\\
NMR/NQR	&		A	&		A	&DD\\
Andreev reflection&	A	&		PA&	DD\\
Pb/c-axis YBCO Josephson junction&	A&			D&		A\\
c-axis BSCCO twist Josephson junction&	DA&		DD&	PA\\
Corner SQUID/Josephson junction&	PA	&	PA	&PA\\
Tricrystal/Tetracrystal Josephson junctions&PA	&PA	&		PA\\
	\end{tabular}
	\end{center}
	\end{table}
\narrowtext
\noindent
T$_{c}$ = 89 K (Ref.~\cite{Kras}), and 2$\Delta_{M}/k_{B}T_{c}$ 
= 5.37  for an overdoped BSCCO  with T$_{c}$ = 80 K (Ref.~\cite{Yu}). 
It 
is apparent that 2$\Delta_{M}/k_{B}T_{c}$ decreases almost linearly 
with  T$_{c}$ in the overdoped range.  Using the fitted curve of 
2$\Delta_{M}/k_{B}T_{c}$  versus T$_{c}$, we 
estimate 2$\Delta_{M}/k_{B}T_{c}$  = 6.04 and  2$\Delta (\theta_{r})$ 
= 
38.6 meV for Y$_{0.9}$Ca$_{0.1}$Ba$_{2}$Cu$_{3}$O$_{7-y}$ with 
T$_{c}$ = 
85.5 K.   Similarly 
we can obtain 2$\Delta_{M}/k_{B}T_{c}$  = 5.66 and 2$\Delta 
(\theta_{r})$ = 
35.1 meV for overdoped BSCCO with T$_{c}$ = 83 K.  INS experiment on 
Y$_{0.9}$Ca$_{0.1}$Ba$_{2}$Cu$_{3}$O$_{7-y}$ (Ref.~\cite{Pai})
indicates $E_{r}^{even}$ = 43 meV $>$ 
2$\Delta (\theta_{r})$, in contradiction with the theoretical model 
\cite{Millis}.  For overdoped BSCCO with T$_{c}$ = 83 K, 
$E_{r}^{odd}$ = 
38 meV (Ref.~\cite{He}) and thus $E_{r}^{even}$ = 45 meV by analogy 
with 
the case of Y$_{0.9}$Ca$_{0.1}$Ba$_{2}$Cu$_{3}$O$_{7-y}$. 
Both $E_{r}^{even}$ and 
$E_{r}^{odd}$ in this overdoped BSCCO  are far larger than 2$\Delta 
(\theta_{r})$, in contradiction with any theoretical models based 
on the d-wave pairing symmetry.

For the underdoped YBa$_{2}$Cu$_{3}$O$_{6.7}$, the even-channel 
magnetic intensity in the 
normal state is a factor of 2.0-2.5 lower than the odd-channel one 
for 
$E$ = 40 meV, which is about 15 meV above the optical magnon gap 
\cite{Fong2000}. This implies that the 
interlayer antiferromagnetic correlation does not influence magnetic 
excitations 
well above the optical magnon gap \cite{Fong2000}.  Since 
the optical magnon gaps for optimally doped 
and overdoped YBCO are much smaller than that for the underdoped 
YBa$_{2}$Cu$_{3}$O$_{6.7}$, one should expect that 
$I^{N}_{odd}(E)/I_{even}^{N}(E)$ $\simeq$ 1 for $E$ $\simeq $ 40 meV 
in optimally doped and overdoped 
\newpage
\noindent
YBCO.  Thus the observed feature 
(c): $I^{S}_{odd}(E_{r})$/$I^{S}_{even}(E_{r})$ $>$$>$ 1 for optimally 
doped and overdoped YBCO can only be explained by an OP that has 
s-wave symmetry and opposite sign in the bonding and antibonding 
electron bands.

For the slightly underdoped YBa$_{2}$Cu$_{3}$O$_{6.92}$, we can also 
deduce the value of $I_{even}^{S}(E_{r})$/$I_{even}^{N}(E_{r})$ using  the measured 
$I_{odd}^{S}(E_{r}) /I_{odd}^{N}(E_{r})$ = 3.6, $I^{S}_{odd}(E_{r})/I_{even}^{S}(E_{r})$ = 5, and the inferred 
$I^{N}_{odd}(E_{r})/I_{even}^{N}(E_{r})$ $\simeq$ 1 (see above 
discussion).  From the measured 
$I^{S}_{odd}(E_{r})/I_{even}^{S}(E_{r})$ = 5, we have 
$I_{even}^{S}(E_{r}) = 0.2 I_{odd}^{S}(E_{r})$ and thus 
$I_{even}^{S}(E_{r})$/$I_{even}^{N}(E_{r})$ $\simeq$ 
0.2$I_{odd}^{S}(E_{r})$/$I_{odd}^{N}(E_{r})$ = 0.72.   This value is 
close to that deduced for the single-layer Tl-2201 ($\simeq$ 1). 
These results consistently suggest that the intraband neutron 
intensity at $E_{r}$ in the superconducting state is close to that in 
the normal state.  
This unique and important feature we 
have identified rules out d-wave order parameter symmetry because 
d-wave symmetry predicts \cite{Brinckmann,Lee} that 
$I_{even}^{S}(E_{r})/I_{even}^{N}(E_{r})$$>$$>$1 for bilayer 
compounds 
and $I^{S}(E_{r})/I^{N}(E_{r})$$>$$>$1 for single-layer compounds.

From Eqs.~\ref{Ne1} and \ref{Ne2}, it is easy to calculate 
$E_{g}$ and $E_{r}$ if
one knows the gap function $\Delta (\theta) $ and the $\theta_{r}$
value. From the measured Fermi surface, one can readily determine
$\theta_{r}$. For example, we find $\theta_{r}$ =
18.4$^{\circ}$ for a slightly underdoped BSCCO from Fig.~3. For
~optimally ~doped cuprates, we get $\theta_{r}$ $\simeq$ 
16.0$^{\circ}$.  If we would use an isotropic s-wave gap function 
$\Delta 
(\theta )$ = 28 meV for a slightly overdoped YBCO, we would have 
$E_{g}$ = 56 meV and $E_{r}$ $>$ 56 meV, which are far larger than 
the 
measured $E_{g}$ = 33 meV and $E_{r}$ = 40 meV 
(Ref.~\cite{Bourges96}).  If we would use a d-wave gap function 
$\Delta 
(\theta )$= 28$\cos 2\theta$ meV, we would have $E_{g}$ = 47.5 meV 
and 
$E_{r}$ $>$ 47.5 meV, which are also far larger than the measured 
values.  
Thus, one cannot quantitatively explain the neutron data in terms of 
d-wave 
and isotropic s-wave  symmetries.

Alternatively, an extended s-wave with eight line nodes ($A_{1g}$
symmetry) is in quantitative agreement with two-thirds of  the 
experiments 
that were designed to test the order-parameter symmetry for 
hole-doped 
cuprates \cite{Zhao}.  The remaining one-third (e.g., 
tricrystal grain-boundary Josephson junction experiments) are 
explained 
qualitatively by Zhao \cite{Zhao} and by Brandow \cite{Brandow}.  In 
Table 1, we compare nearly all the experiments used to test the OP 
symmetry with the extended s-wave, d-wave, and isotropic s-wave 
models.  
In both extended s-wave and isotropic s-wave models, the order 
parameters are assumed to have opposite sign in the bonding and 
antibonding 
electron bands formed within the Cu$_{2}$O$_{4}$ bilayers 
\cite{Mazin}.  The 
detailed comparisons with other experiments are made in 
Ref.~\cite{Zhao}.  From Table 1, one can see that the neutron data 
alone provide a definitive answer to the intrinsic, bulk OP symmetry 
because INS is a bulk, phase and angle sensitive technique.  Other 
bulk and non-phase sensitive experiments provide complementary 
support 
to the present conclusions.  The phase and surface sensitive 
experiments cannot definitively 
determine the intrinsic bulk OP symmetry because the surface OP might 
be different from the bulk 
one \cite{Zhao}.

For a slightly overdoped YBCO, more than six independent experiments 
consistently suggest that \cite{Zhao} the gap function is $\Delta 
(\theta ) = 24.5(\cos 4\theta + 0.225)$ meV.  Substituting 
$\theta_{r}$ = 16$^{\circ}$ into the gap function, we get $\Delta 
(\theta_{r})$ = 16.3 meV and thus $E_{g}$ = 32.6 meV, in quantitative 
agreement with the measured one (32-33 meV).

 In our model, the position of the extended saddle point 
along the $\theta_{r}$ direction is located at $E_{r}-E_{g}/2$ in the 
superconducting state (see Eqs.~\ref{Ne1} and \ref{Ne2}).  For 
optimally doped YBCO with $E_{r}$ = 41 meV and $E_{g}$ = 32 meV, we 
find that the saddle point in the superconducting state is located at 
an energy of 25 meV below the Fermi level.  This is in agreement with 
the ARPES studies \cite{King} which suggest that the Fermi level in 
the superconducting state for optimally doped cuprates is $\leq$ 30 
meV above the extended saddle points that have the same energy over a 
large momentum space \cite{Dessau}.  Further, electronic Raman 
scattering spectra in YBa$_{2}$Cu$_{4}$O$_{8}$ have been used to 
determine the energy of the extended saddle points more accurately 
\cite{Sherman}.  At 10 K (well below $T_{c}$), the energy of the 
extended saddle points is found to be 24.3 meV below the Fermi level, 
in excellent agreement with the result predicted by our model from 
the 
INS data ($\simeq$ 25 meV).

We have identified the unique and important 
feature: $I^{S}(E_{r})$/$I^{N}(E_{r})$ $\simeq$ 1.0 for the 
intralayer 
and intraband magnetic scattering.  This feature unambiguously rules 
out 
the d-wave OP symmetry.  The unambiguous determination of the 
intrinsic extended s-wave pairing symmetry for hole-doped cuprates 
places strong constraints on the pairing mechanism for 
high-temperature superconductivity.  A recent calculation  suggests 
that high energy Cu-O charge fluctuations can 
lead to an attractive interaction between conduction electrons and 
the 
pairing symmetry may be of extended s-wave ($A_{1g}$) \cite{Miklos}.

Indeed, precise thermal-difference reflectance 
spectra of several cuprate superconductors ($T_{c}$ =105-120 K) 
exhibit 
pronounced features at photon energies of about 2.0 eV, which may be 
related to the Cu-O charge fluctuations (Ref.~\cite{Little}).  These 
features can be well described within 
Eliashberg theory with an electron-boson coupling constant 
$\lambda_{ch}$ of 
about 0.40.  In order to explain a superconducting transition 
temperature of 105-120 K, 
the authors simulated \cite{Little} an electron-phonon coupling 
feature at 50 meV with the coupling constant $\lambda_{ph}$ of about 
1.0.  This Eliashberg model's simulated electron-phonon coupling 
agrees 
well with the results obtained from  first principle calculations 
\cite{Kra}. 

The contribution of this 2 eV component can be 
equivalent to a {\em negative} Coulomb pseudopotential $\mu^{*}$ 
within Eliashberg theory 
\cite{Zhao1}.   By using a realistic electron-phonon coupling 
spectral 
weight deduced from tunneling spectra and a $\mu^{*}$ = -0.15, and 
taking into a polaronic effect, Zhao 
{\em et al. } \cite{Zhao1} are able to explain the negligible isotope 
effect 
on $T_{c}$, and the magnitudes of $T_{c}$ and the superconducting gap 
for  
optimally doped 90 K superconductors.

In summary, we have identified several important features in the 
neutron scattering data of 
cuprates, which are difficult to be explained in terms of d-wave and 
isotropic s-wave 
order parameters.  Alternatively, we show that the neutron data are 
in 
quantitative agreement with an order parameter that has an extended 
s-wave ($A_{1g}$) symmetry and opposite sign in 
the bonding and antibonding electron bands formed within the 
Cu$_{2}$O$_{4}$ bilayers.   This $A_{1g}$ pairing 
symmetry may be compatible with a charge fluctuation mediated pairing 
mechanism.  High-temperature superconductivity in cuprates may be due 
to 
the combination of strong electron-phonon coupling and substantial 
electron-charge-fluctuation coupling \cite{Little,Zhao1}. 
~\\
~\\
$^{*}$Correspondence should be addressed to gzhao2@calstatela.edu

\end{document}